# Observations of near-perfect nonclassical correlation using coherent light

Sangbae Kim and Byoung S. Ham*

Center for Photon Information Processing, School of Electrical Engineering and Computer Science, Gwangju Institute of Science and Technology
123 Chumdangwagi-ro, Buk-gu, Gwangju 61005, South Korea

*bham@gist.ac.kr

(submitted on May 04, 2021)

**Abstract:**

Complementarity theory is the essence of the Copenhagen interpretation. Since the Hanbury Brown and Twiss experiments, the particle nature of photons has been intensively studied for various quantum phenomena such as anticorrelation and Bell inequality violation in terms of two-photon correlation. Regarding the fundamental question on these quantum features, however, no clear answer exists for how to generate such an entanglement photon pair and what causes the maximum correlation between them. Here, we experimentally demonstrate the physics of anticorrelation on a beam splitter using sub-Poisson distributed coherent photons, where a particular photon number is post-selected using a multiphoton resolving coincidence measurement technique. According to Born rule regarding self-interference in an interferometric scheme, a photon does not interact with others, but can interfere by itself. This is the heart of anticorrelation, where a particular phase relation between paired photons is unveiled for anticorrelation, satisfying the complementarity theory of quantum mechanics.

**Introduction**

Quantum features such as Bell inequality violation[1-6] and anticorrelation[7-10] between two-mode quantum particles or engineered individual photon pairs have been intensively studied over the last few decades to understand the quantum nature of entanglement beyond the classical limits of individuality and separability[11]. The physical understanding of quantum entanglement based on the particle nature of photons governed by the energy-time relation via the uncertainty principle has been applied for various quantum information technologies. Although the complementarity theory of quantum mechanics is normally limited to the microscopic world of conjugate entities satisfying the uncertainty principle[11,12], the Schrodinger cat[13-15] as a macroscopic quantum feature should also be confined by the same physics. Here, we report the first observation of on-demand quantum correlation using sub-Poisson distributed coherent photons obtained from an attenuated continuous wave (cw) laser. Since the wave nature of such sub-Poisson distributed photons relies on the intrinsic properties of the coherent light source, the degree of quantum correlation among photon pairs can be deterministically manipulated not only for microscopic entities, but also for a macroscopic Schrodinger cat. This paper opens the door to on-demand quantum entanglement generation from any light source.

Since the seminal paper of EPR[16] in 1935, entanglement representing the maximum correlation between two individual photons or atoms has been demonstrated via the second order intensity correlation $g^{(2)}(\tau)$, where $\tau$ is the time delay between two particles for coincidence detection in an interference scheme[7-11]. Bell inequality violation is for a nonlocal quantum feature in a noninterference scheme[1-6], while anticorrelation, the so-called Hong-Ou-Mandel (HOM) dip, is for the same quantum feature in an interference scheme[7-10]. If a single photon as a Fock state is considered[17], a nondefinite phase of the Fock state is assigned due to the energy-time relation of the uncertainty principle. Thus, all quantum features denoted by quantum operators in a space-time domain have no definite phase relation at all. The wave nature of a single photon has been mostly forgotten in the areas of quantum information, even though a definite phase relation between paired photons is an essential requirement[7-10]. In that sense, the terminology of coherence has been used differently depending on the context[18] ever since the temporal correlation between two individual photons was demonstrated by Hanbury Brown and Twiss[19].

Recently, a completely different approach has been pursued for the study of anticorrelation to unveil the quantum nature of photon bunching on a beam splitter (BS)[20]. Such a pure coherence optics-based interpretation for a quantum feature has been successful to describe a HOM dip. As already known, experimental results of Young's double slits or a Mach-Zehnder interferometer (MZI) show the same interference fringe whether the inputs are single photons or coherence light[21-23]. This is based on Born rule of quantum mechanics regarding superposition of a single photon, where single photons do not interfere or interact with others, but with itself[24]. The wave-particle duality has also been discussed for the interference scheme, where the definition of particle-like or wave-like photons is based on coherence[25-27]. As a result, there is no fundamental difference between wave and particle approach[28]. In the wave approach, self-interference in Young's double-slit experiments or a Mach-Zehnder interferometer (MZI) is the bedrock of all quantum phenomena in an interferometric scheme regardless of photon number or photon characteristics whether it is viewed as a particle or a wave. Here, we experimentally demonstrate near-perfect nonclassical correlation using sub-Poisson distributed coherent photons obtained from attenuated laser light to prove recently proposed theory[20]. This is the first step toward on-demand quantum feature generation based on the wave nature of photons under the Copenhagen interpretation of quantum mechanics.

**Experimental setup**

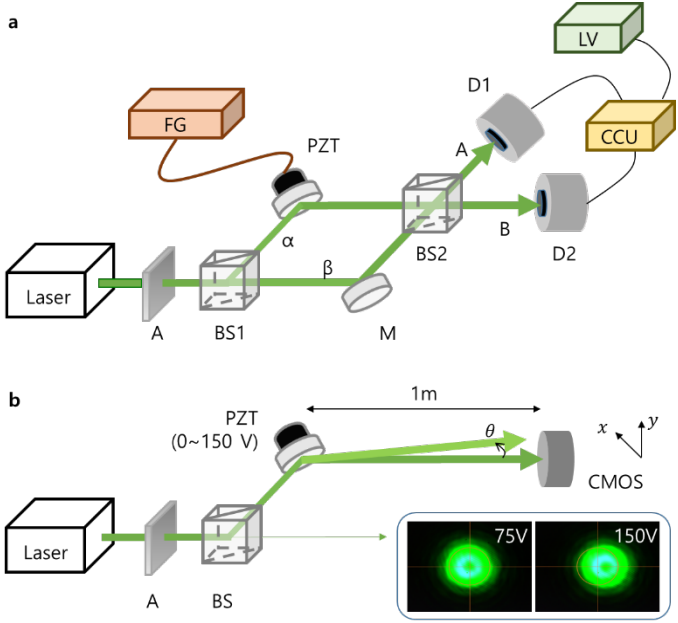

**Fig. 1. Schematic of quantum feature generation using post-selected coherent states. a** An interferometric scheme. **b** A one-axis PZT scan-induced decoherence for **a**. A: attenuator, BS: nonpolarizing 50/50/beam splitter, M: mirror. D: a single photon detector; CCU: coincidence counting module; LV: Labview. PZT: Piezo-electric transducer; CMOS: 2D image sensor.

Figure 1 shows schematic of the coherent photon-based anticorrelation as proposed in ref. 20, where the photon numbers are post-selected via coincidence measurements. The laser is cw at a wavelength of 532 nm (Coherent, Verdi-V10), whose spectral bandwidth is 5 MHz, resulting in 0.2 µs of coherence time or 60 m of coherence length. The power stability is ~1 %. With proper attenuation using neural density filters (OD~10), a sub-Poisson distributed coherent photon stream is obtained, where the mean photon number is $\langle n \rangle \sim 0.04$ (see Figs. S1-S3 of the supplementary Information). Here, the sub-Poisson distributed coherent photons have the same bandwidth as the original coherent laser light regardless of the mean photon number. This is the essence of the self-interference governed by Born rule[28]. Regarding the MZI in Fig. 1a, the first BS (BS1) plays a role of phase shift of $\pi/2$ between the reflected and transmitted photon pair[29]. Thus, the paired photons experience a destructive interference on the second BS (BS2), resulting in photon bunching either A or B depending on the PZT (Thorlabs, POLARIS-K2S3P)-induced difference phase of $\pi$ or 0, respectively. Regarding the PZT control, however, only one axis out of two-axis mirror control unit (Thorlabs, POLARIS-K2S3P) via a PZT controller (Thorlabs, MDT693B) is scanned to induce intentional decoherence as shown in Fig. 1b. The extreme decoherence results in the classical bound as a reference (see Fig. 2 and Methods). For this, the PZT is set to make both beams overlap at the center of the scan range at 75 V (see the Inset of Fig. 1b). This on-demand decoherence technique gives us a clear understanding of the wave nature for conventional measurement systems based on wide-bandwidth photons as well as practical advantages in determining the two-photon correlation limit.

Unlike conventional anticorrelation experiments based on spontaneous parametric down conversion processes (SPDC), whose generated photon pairs have a random phase among them[7-10], all incident photons of Fig. 1 have the same phase within the narrow bandwidth of 5 MHz. Each MZI output signal $(A; B)$ is detected by a single photon counting module (SPCM; Excelitas AQRH-SPCM-15: $D_1; D_2$). The time resolving of the SPDM is 350 ps at $\lambda = 825$ nm with a dead time of 22 ns, and its dark count rate is measured for $27 \pm 5$ counts per second (cps). Both electrically converted single photon signals from SPCMs are fed into a coincidence counting unit (CCU; Altera DE2) for coincidence detection measurements.

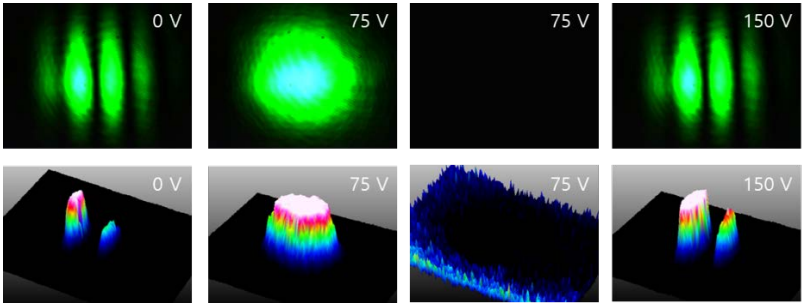

**Fig. 2. One-axis PZT controlled coherence modification (see Fig. 1b).** (Top) 2D images captured on CMOS. (Bottom) 3D analyses for Top images.

Figure 2 shows experimental demonstrations of decoherence control by using one-axis controlled PZT (Thorlabs, POLARIS-K2S3P) in Fig. 1b. For this, the cw power is set at 0.3 µW. The top panels of Fig. 2 represent 2D images captured on a CMOS camera (Thorlabs, DC3241M), while the bottom panels show their corresponding 3D analyses by a corresponding manufacturer supplied software. The second and third columns at 75 V of PZT are for constructive and destructive interferences, respectively, by a fine path control of the MZI. At both ends of 0 V and 150 V, multiple fringes result in the nearly path-length independent intensity outputs, corresponding to the incoherence feature of interferometry.

**Results**

*Preparation of a single photon stream*

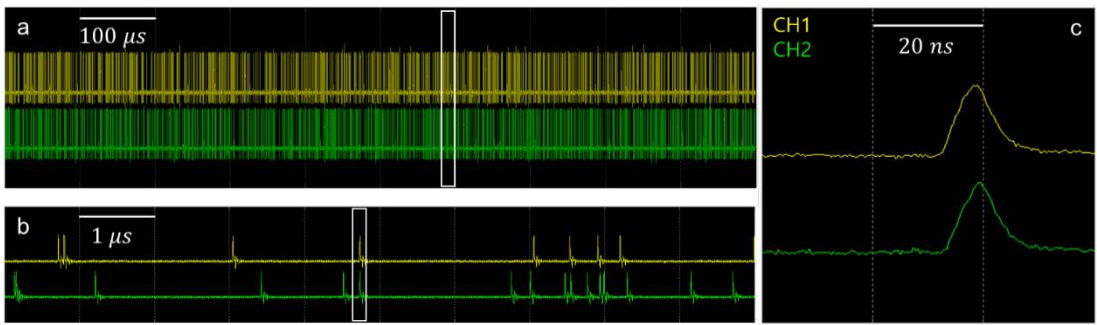

**Fig. 3. Sub-Poisson distributed photon characteristics observed by a fast digital oscilloscope. a** Photon streams detected by D1 (yellow) and D2 (green) in Fig. 1. **b** Expansion of the box in **a**. **c** Expansion of the box in **b** for a coincidence photon pair.

To provide sub-Poisson distributed photons, we analyze the attenuated photon statistics from the 532 nm laser using both SPCMs and a high-speed digital oscilloscope (Yokogawa DL9040; 500 MHz). For these measurements, the photon statistics is pursued for a noninterference regime by simply removing BS2 in Fig. 1.

First, we perform photon counting measurements, analyze the obtained data, and compare them with the Poisson statistics for dozens of different OD values from the ND filter combinations (see Methods and Section A of the Supplementary Information). Second, the same measurements are performed using the oscilloscope to visualize single photon streams, to obtain separate photon statistics, and to compare it with the CCU-based ones (see Section B of the Supplementary Information). With these two different measurement techniques, we confirm that the Poisson statistics works well for the attenuated coherent laser and the present coincidence detection measurements under the sub-Poisson photon statistics.

Figure 3 shows a captured data from the digital oscilloscope via SPCMs. The top two rows in Fig 3a show photon streams detected by both SPCMs of D1 (yellow) and D2 (green) in Fig. 1, respectively. Figure 3b is an expansion of Fig. 3a for the box. Figure 3c is an expansion of Fig. 3b for a bunched photon case. The occurrence rate of such doubly bunched photons in Fig. 3c to the single photons in Fig. 3b is ~1 % for the present observations, where details rely on the mean photon number $\langle n \rangle$.

*Experimental demonstrations of nonclassical features via coincidence measurements*

For Fig. 4, the PZT in Fig. 1 is continuously and repeatedly scanned for the same region via a series of saw tooth ramps generated by a function generator (Tecktronics AFG3102; 0~150V), where the center position is fixed for a nearly perfect overlap on BS2, resulting in a symmetric decoherence feature as shown in Fig. 2. For the optimum counting rate with respect to the PZT scanning, accumulation time of 0.1 s is set for CCU. Considering 10 fringes per scan range in Fig. 4a, 200 data points are allocated to each fringe. This corresponds to 0.01 π radians (1.8 degrees) per data point, which is good enough for high resolution measurements.

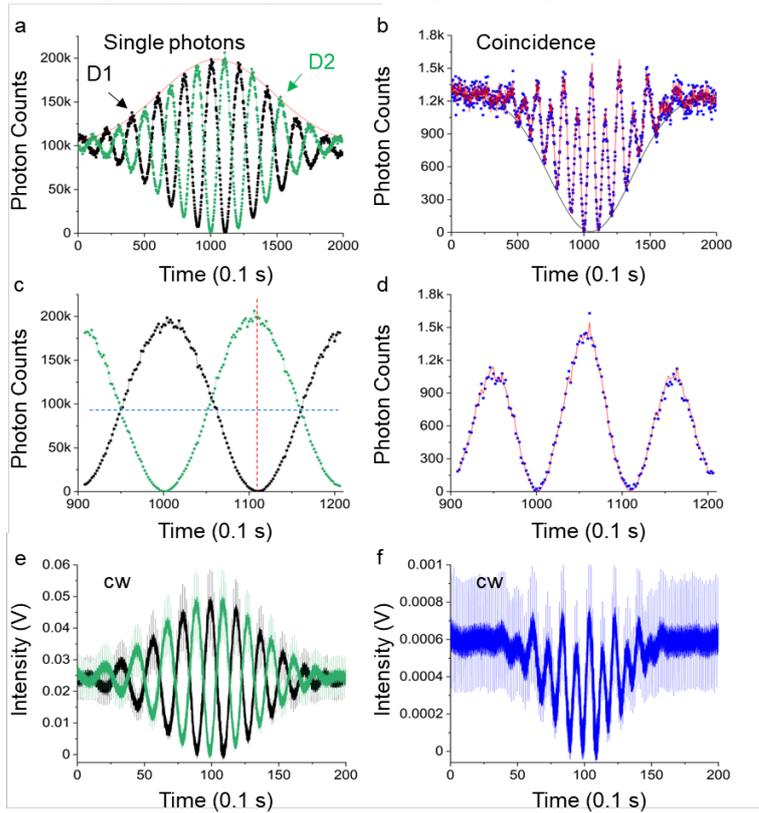

**Fig. 4. Experimental results for Fig. 1. a** Single photon counting rates measured by D1 (black dots) and D2 (green dots). **b** Measured coincidence detection for **a**. **c** and **d** Respective expansions of **a** and **b**. **e** Conventional cw MZIs for D1 (black) and D2 (green), where SPCMs are replaced by avalanche photon diodes. **f** Products of two data points in **e** directly performed on the oscilloscope. The cw laser power used for **e** and **f** is 10 μW.

Figure 4a is for individual single photon measurements for both outputs A and B in Fig. 1. As shown in Fig.

4a, the one-axis PZT scan-induced symmetric decoherence is demonstrated successfully, where both ends of the fringes represent incoherence, i.e., the classical bound. The red curve is a Gaussian fit for the measured fringes. According to the equality between visibility and coherence, the asymmetric PZT scan technique in Fig. 1 is confirmed for the intentional decoherence purpose. The measured visibility at near center position is 99.9 % from the minimum counts of 105 with respect to the maximum counts of 200,000. This measured visibility is unprecedented and the highest value in quantum correlation measurements so far. Here it should be noted that the classical bound of visibility is 70.7% according to the Bell inequality[1-6]. Thus, Fig. 4a experimentally demonstrates a quantum feature in an interferometric scheme for the attenuated coherent photons. The deterministic fringe pattern in Fig. 4a does not violate the randomness of path selection by each single photon in an MZI via Born rule of self-interference in quantum mechanics. The superposition principle should be limited by a single photon, where the fringe in Fig. 4a is only due to the self-interference of each photon.

Figure 4b shows the coincidence measurements for Fig. 1. The blue dots are the experimental data measured by CCU. The black curve is a best-fit curve from the product of each data set in Fig. 4a, which is nearly perfectly coincides with the measurements. The product is for the definition of AND gate operation of CCU. Such a relation between the CCU logic and coincidence measurements never been demonstrated before in conventional SPDC-based HOM dips. The reason is inherent due to the random frequency detuning swapping between signal and idler photons, resulting in random output selections between A and B. The minimum photon counting rate in Fig. 4b is 2 counts per 0.1 s. According to anticorrelation[20], the lower bound of classical physics in the intensity correlation $g^{(2)}(0)$ is 0.5. Reminding that the intentional set up for incoherence optics with asymmetric PZT scanning in Fig. 1b and Fig. 2, the maximum counting rate in Fig. 4b indicates the classical lower bound. According to the product of individual data in Fig. 4a, all crossing points result in maxima in Fig. 4b, indicating incoherence or classical particles (discussed in Fig. 5). Thus, all measurement data points below this reference at ~1,200 counts per 0.1 s represent a quantum feature. This quantum feature also shows a fringe due to the relative phase relation of $sin^2(\varphi)$ as theorized in ref. 20. Thus, Fig. 4b clearly demonstrates new physics of the wave nature of photons regarding anticorrelation or HOM dip based on coherent photons. This deterministic coherent light-based quantum feature is new and unprecedented.

Figures 4c and 4d are expansions of Figs. 4a and 4b, respectively. As shown in Fig. 4c, both fringes of single photon measurements are not perfectly aligned (see the vertical red dashed line). Such misalignment is due to the imperfect overlap on BS2, intensity fluctuation of the laser, or MZI phase fluctuation by air turbulence. As a result, the crossing points contributed to maxima in Figs. 4b and 4d are also misaligned (see the horizontal blue dashed line). As analyzed in Fig. S7 of the Supplementary Information, there is just one per cent misalignment error in Fig. 4 (discussed in Fig. 5).

Figures 4e and 4f are for the corresponding cw cases of Figs. 4a and 4b, respectively. The plots of Figs. 4e and 4f are from the stored data in the fast digital oscilloscope. For this, the SPCMs in Fig. 1 is replaced by avalanche photon diodes (Thorlabs, APD 110A) and connected directly to the oscilloscope (YOKOGAWA DL9040). Both fringes in Fig. 4e are from both channels of the oscilloscope, while Fig. 4f is generated from the product function of the oscilloscope for Fig. 4e. Because the physics of the present anticorrelation is based on the wave nature of photons, there should be no difference between coincidence measurements of single photons in Fig. 4b and conventional wave optics of MZI outputs in Fig. 4f. According to Born rule of quantum mechanics regarding self-interference, an MZI corresponding to a two-slit case actually confines the path superposition to a single photon case, where the coincidence measurement is simply a product of each single photon measurement[22-24]. This rule is also applied to the conventional SPDC-based anticorrelation, where the HOM dip is caused by photon bunching due to destructive interference in an MZI of Figs. 4b and 4f[20]. As already demonstrated by Born rule tests[28], Fig. 4 also demonstrates the fact that there is no difference between the particle (single) and wave (cw) natures of photons in an interferometric scheme. This experimental demonstration for coincidence measurement in Fig. 4 is unprecedented.

Figure 5 provides analyses for Figs. 4a and 4b with perfect overlap for the observed nonclassical features using coherent photons. For this, all data from D1 in Fig. 4a is moved by two pixel points, resulting in nearly perfect coincidence measurements in Fig. 5b as expected by the logic gate operation discussed in Figs. 4b and 4d. Figures 5c and 5d are numerical simulations corresponding to Figs. 5a and 5b, respectively. The coincidence rate in Fig. 5d is normalized, where the all maxima coincide each other as expected. Here, the meaning of the same maxima is self-evident by definition of product between D1 and D2 for AND operation. Thus, the maxima in Fig. 5b correspond to the crossing points in Fig. 5a as shown. This crossing point of both channel-based fringes indicate no coherence in the interference of Fig. 1. In other words, the maxima in Fig. 5b is for incoherent light as a classical reference, which is designed by the present asymmetric PZT scanning method. Except maxima in Figs. 5b and 5d, all data points are nonclassical by the definition of incoherence as well as intensity correlation of quantum mechanics. Thus, the observed quantum correlation in Fig. 4b is strongly related with coherence (relative phase) between paired photons, resulting in on-demand control of quantum features. The maximum correlation (zero photon count) occurs at the relative phase $\varphi \in \{0, \pi\} = n\pi$, which are the phase bases of an MZI. This is the heart of the present demonstrations of deterministic nonlocal correlation for future quantum technologies.

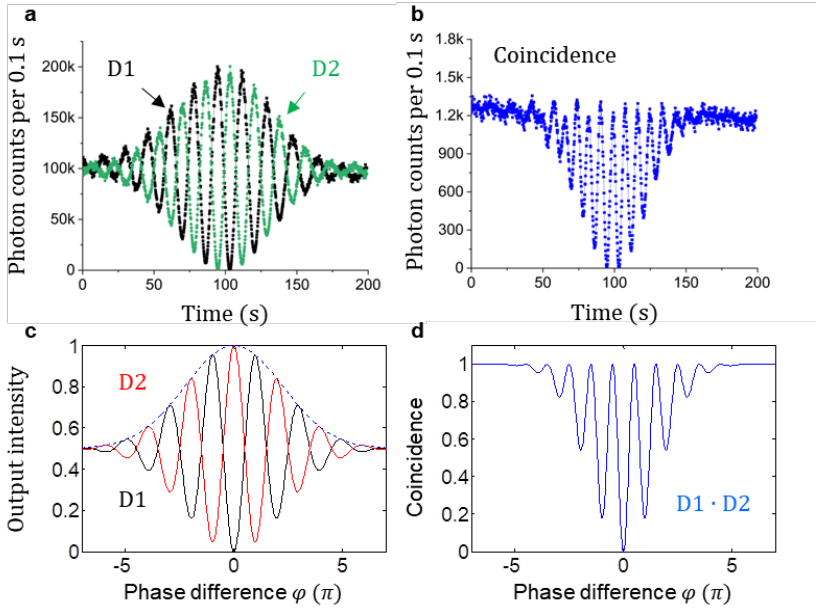

**Fig. 5. Analysis of quantum correlation based on coherent photons. a** and **b** All D1 data in Fig. 4a are shifted by two pixel points. **c** and **d** Numerical simulations for **a** and **b**, respectively. The dotted curve in **c** is a Gaussian function applied. The coincidence value in **d** is normalized.

**Conclusion**

We experimentally demonstrated the novel feature of on-demand two-photon correlation corresponding to a HOM dip or anticorrelation, whose manipulation is not probabilistic but deterministic with difference phase between paired photons. For experimental demonstrations of this deterministic quantum feature, we used sub-Poisson distributed photons obtained from an attenuated laser. We tested the photon statistics using a fast digital oscilloscope and compared it with that obtained from coincidence detection measurements. For the demonstrations of anticorrelation using coherent photons, a simple MZI scheme was adapted, where the first BS plays a role of a $\pi/2$ phase shift between coincident photon pairs. To induce an artificial incoherence effect as a classical bound, an asymmetric PZT scanning method was developed, otherwise a long coherence time was kept for the coincidence detection. For coincidence detection measurements between two output ports of the MZI, we confirmed the new physics of anticorrelation proposed in ref. 20 by demonstrating near perfect visibilities in each channel as well as a wavelength dependent intensity correlation fringe for coincidence measurements. To support the experimental data, a simple Gaussian adjusted MZI simulations were compared, resulting in a near perfect coincidence with the experimental data. Thus, the observed coincidence measurements based on coherent photons demonstrated the deterministic control of anticorrelation based on the wave nature of photons without violating quantum mechanics. Most of all, the present demonstration witnesses the new interpretation of quantum mechanics, satisfying Copenhagen interpretation of the wave-particle duality as well as self-interference of Born rule. By replacing the single photons with a cw laser, the thumb rule of no distinction between quantum particles and classical waves in an interferometric scheme was experimentally demonstrated for the indirect Born rule test.

**Methods**

A standard Mach Zehnder interference (MZI) test is performed using an unattenuated cw laser of Coherent Verdi V10 at $\lambda = 532$ nm, whose spectral bandwidth is 5 MHz. For the sub-Poissonian distributed single photon generations, conventional coincidence measurement technique is applied. To confirm the sub-Poissonian distributed photon statistics, photon visualization is additionally conducted using a fast digital oscilloscope (YOKOGAWA DL9040; 500 MHz; 400 ps resolving time for two-channel access). For this, the SPCMs are directly connected to the oscilloscope. The size of the MZI in Fig. 1 is $\sim 10 \times 10 cm^2$, enclosed by a cartoon box to minimize air fluctuation-caused phase noise. Because the bandwidth-caused coherence length of the laser is long enough to be 60 m, the PZT-caused MZI interference fringes observed in Figs. 2-5 are well defined within the PZT length variations of a few microns. For conventional cw MZI tests in Fig. 4e and 4f, the SPCM for D2 is replaced by an avalanche photon diode (Thorlabs, APD 110A) and directly connected to the oscilloscope.

1. Dark count measurements for a single photon detection module

In a dark room condition enclosed by a black cartoon box, both single photon counting modules (Excelitas AQRH-SPCM-15) are tested for the dark count rate of $27 \pm 5$ (counts/s) without input photons, which is satisfied with the manufacture specification of 50 (count/s).

2. Attenuated 532 nm laser

The attenuated light source is Coherent cw laser (Verdi V10). The output intensity is stabilized at ~1 % variation for the fixed output power of 10 mW. For the cw MZI experiments in Fig. 4, OD 3 of ND filters are placed before MZI, resulting in 10 μW. For the single photon MZI experiments in Figs. 2~5, another OD 10 is added, resulting in 1 fW power and a mean photon number of $\langle n \rangle \sim 0.04$.

3. Counting method for single and bunched photons by a coincidence counting module

The SPCM-generated electrical pulses are sent to the coincidence counting module (CCU: DE2 FPGA, Altera). The pulse duration of each electrical signal from SPCM for single photon detection is ~10 ns. Both single and coincidence counting numbers are counted in parallel by DE2 for 100 ms acquisition time for each data point, otherwise specified. The measured data is transferred to Labview via RS232 cable for a coincidence_rs232(4_5).vi application program in real time. For single and coincidence measurements in Fig. 4, the measured photon number is compared with that by an oscilloscope-based visualization technique below. The acquisition time of 100 ms for the CCU was set for an optimum number between the MZI fringe duration and the counted coincidence detections. Additionally, photon statistics for single photons, doubly bunched, and triply bunched photons, a conventional photon resolving measurement technique are used. The results are shown in Fig. S2 of the Supplementary Information. In this data, higher-order bunched photon ratio to the doubly bunched photons is less than 1 %. The mean photon number is measured at $\langle n \rangle = 0.04$.

4. Counting method for single and bunched photons by a fast digital oscilloscope

For this, SPCMs were directly connected to the oscilloscope (YOKOGAWA; DL9040). Channel 1 of the oscilloscope is for D1, while Channel 2 is for D2. Due to the storage limitations of the oscilloscope for a displaced wave form on a screen, a stored 1 ms data length is analyzed for both sub-Poissonian photon distribution and coincidence detections. The results are also compared with the SPCM-based photon statistics. To count single photons in each data set, a homemade MATLAB program is used (see code in Fig. S6 of the Supplementary Information). For 10 measured data samples (not shown), the counted numbers for single photons are tabulated in Table S3 of the Supplementary Information.

5. Asymmetric PZT scanning

The coherence-length modification technique in Fig. 1b is based on the linearly increased phase shift via walk-off cross section between two lights on BS2 in Fig. 1, where such misalignment is conducted by an asymmetric control of a PZT kinematic mirror (Thorlabs, POLARIS-K2S3P) via a PZT controller (Thorlabs, MDT693B). For this, a PZT scanning applies to one of two mirror control knobs, where the midpoint of the scanning range is set to be maximally coherent at 75 V, whose PZT voltage resolution is 1.5 mV.

**Data availability**

Data for figures are available upon reasonable requests.

**Acknowledgments** This work was supported by GIST via GRI 2020.

**Author contributions** B.S.H. conceived the research on both theory and experiments, did numerical simulations, data analysis, and wrote the manuscript; S.K performed all experiments and analyzed the related data.

**Correspondence and request of materials** should be addressed to BSH (email: bham@gist.ac.kr).

**Competing interests** The author declares no competing (both financial and non-financial) interests.

**Supplementary information** is available in the online version of the paper.